\documentclass[aps,twocolumn,prb,10pt,amsmath,amssymb,floatfix]{revtex4-2}

\usepackage[urlcolor=blue, colorlinks=true, citecolor=blue, linkcolor=blue]{hyperref}
\usepackage[dvips]{graphics}
\usepackage{bbm,mathtools,color,dcolumn,multirow,booktabs}

\newcommand{\vk}{\vec k}

\renewcommand{\vr}{\vec r}

\newcommand{\vR}{\vec R}

\newcommand{\ZZ}{\mathbb{Z}}

\newcommand{\tr}{{\rm tr}\, }

\newcommand{\id}{\textbf{1}}

\renewcommand{\vec}[1]{\mathbf{#1}}

\newcommand\rotn{\mathcal{C}}
\newcommand\mirror{\mathcal{M}}
\newcommand{\trev}{\mathcal{T}}

\newcommand{\cmplx}{\mathbb{C}}
\newcommand{\intg}{\mathbb{Z}}
\newcommand{\real}{\mathbb{R}}
\newcommand\Gr{\text{Gr}}

\newcommand\G\Gamma
\newcommand\K{ {\mathrm{K}} }
\newcommand\M{ {\mathrm{M}} }

\newcommand\C{ \mathrm{C} }
\newcommand\D{ \mathrm{D} }
\newcommand\e{ \mathrm{e} }

\newcommand\hsp{\mathcal{H}}

\newcommand{\nullv}{\mathbf{0}}

\newcommand{\ve}{\varepsilon}

\newcommand{\dg}{{\dagger}}

\newcommand{\abs}[1]{\left| #1 \right|}

\renewcommand\hom[2]{\pi_{#1}\!\left[#2\right]}
\newcommand{\sgn}[1]{\text{sgn}\left( #1 \right)}

\graphicspath{{./}{./figures/}}

\newcommand\wpos{\xi}
\newcommand\I{ \mathrm{i} }

\begin{document}

\title{On the band topology of the breathing kagome lattice}
\author{Clara K.\ Geschner}
\author{Adam Yanis Chaou}
\author{Vatsal Dwivedi}
\author{Piet W.\ Brouwer}
\affiliation{Dahlem Center for Complex Quantum Systems and Institut f\"ur Physik, Freie Universit\"at Berlin, Arnimallee 14, D-14195 Berlin, Germany}

\begin{abstract}
A two-dimensional second-order topological insulator exhibits topologically protected zero-energy states at its corners. In the literature, the breathing kagome lattice with nearest-neighbor hopping is often mentioned as an example of a two-dimensional second-order topological insulator. Here we show by explicit construction that the corner states of the breathing kagome lattice can be removed by a continuous change of the hopping parameters, without breaking any of the model's symmetries, without closing bulk and boundary gaps, and without introducing hopping terms not present in the original model. Furthermore, we topologically classify all three-band lattice models with the same crystalline symmetries as the breathing kagome lattice and show that though none of the phases have protected zero-energy corner states, some of the phases are obstructed atomic limits which exhibit a filling anomaly. 
\end{abstract}

\maketitle

\emph{Introduction. ---} 
A two-dimensional insulator or superconductor is said to possess second-order topology if it has anomalous zero-energy states localized at its corners, which cannot be removed without closing the bulk gap \cite{benalcazar2017, benalcazar2017b, parameswaran2017, peng2017, song2017, schindler2018, langbehn2017, geier2018, fang2019, khalaf2018, khalaf2018b, trifunovic2019, trifunovic2020}. Such anomalous corner states are a manifestation of crystalline topology, \emph{i.e.}, topology that is protected by crystalline symmetries, such as rotation or mirror symmetries \cite{trifunovic2020,fu2007b,fu2007,fu2011,turner2012,chiu2013,morimoto2013,shiozaki2014,ando2015,trifunovic2017}. These corner states are pinned to zero energy by an antisymmetry of the Bloch Hamiltonian, such as a particle-hole or sublattice symmetry \cite{schnyder2008, schnyder2009, kitaev2009}.

Soon after the publication of the first articles on second-order topological insulators, Ezawa \cite{ezawa2018} proposed the nearest-neighbor tight-binding model on the \emph{breathing kagome lattice} \cite{schaffer2017} --- a two-dimensional lattice of corner-sharing triangles with different inter- and intra unit cell hoppings --- as a realization of a second order topological phase. While this system indeed hosts zero-energy modes localized at the corners, their interpretation as a signature of second-order topology is at odds with the observation that the bulk spectrum of the breathing kagome is not symmetric around zero energy, precluding a particle-hole or sublattice symmetry required for second-order topology in two dimensions. Moreover, the understanding of anomalous corner modes as a domain wall between boundary masses of opposite sign \cite{schindler2018, langbehn2017, geier2018, fang2019} is inconsistent with second-order topology protected by $\C_3$ rotation symmetry. Nonetheless, the claim that the breathing kagome lattice realizes a second order topological phase has been widely quoted in the literature, see, {\em e.g.}, Refs. \cite{ezawa2018c,xue2019,ni2019,kempkes2019robust,li2019higher,elhassan2019,xie2021,wang2021higher,yatsugi2023higher}, with Ezawa's first paper \cite{ezawa2018} cited over 700 times.

That the zero-energy corner states of the breathing kagome lack topological protection has been previously pointed out in two theoretical studies \cite{vanmiert2020,herrera2022}, which constructed explicit perturbations to the model that move the corner states away from zero energy while respecting all relevant symmetries and without closing the bulk gap. These constructions, however, require symmetry-allowed hoppings beyond the nearest-neighbor ones. In this article, we demonstrate that the corner states of the breathing kagome lattice can be removed using symmetry-allowed changes of nearest-neighbor hoppings only, without the addition of hopping elements that are not contained in the original model. Our conclusion, therefore, reinforces that of Refs.~\cite{vanmiert2020,herrera2022} that the zero-energy corner modes of breathing kagome are accidental and not a signature of second-order topology.

While the breathing kagome lattice does not possess second order topology, it nevertheless exhibits topologically nontrivial {\em obstructed atomic limits} \cite{bradlyn2017,xu2024,benalcazar2019} --- topological crystalline phases which are continuously deformable to a model with strictly localized eigenstates, but where the crystalline symmetry prohibits deformations to a trivial reference phase. Such phases may exhibit a \emph{filling anomaly} \cite{fu2011,vanmiert2018b,benalcazar2019,watanabe2020,trifunovic2020b,fang2021} --- the existence of a net charge in a lattice that obeys the crystalline symmetries for a completely filled valence band. We further perform an exhaustive classification of three-band lattice models on the breathing kagome which respect the relevant crystalline symmetries and show which of the topological phases exhibit a filling anomaly. (We also verify that none of the phases exhibit second order topology, as was to be expected given the absence of a spectral antisymmetry.) We derive these results both in presence and absence of a ``tripartite symmetry'' \cite{kempkes2019robust,ni2019,li2020,li2021}, which forbids matrix elements between sites belonging to the same sublattice. 

\begin{figure}[t]
\centering
\includegraphics[width=\columnwidth]{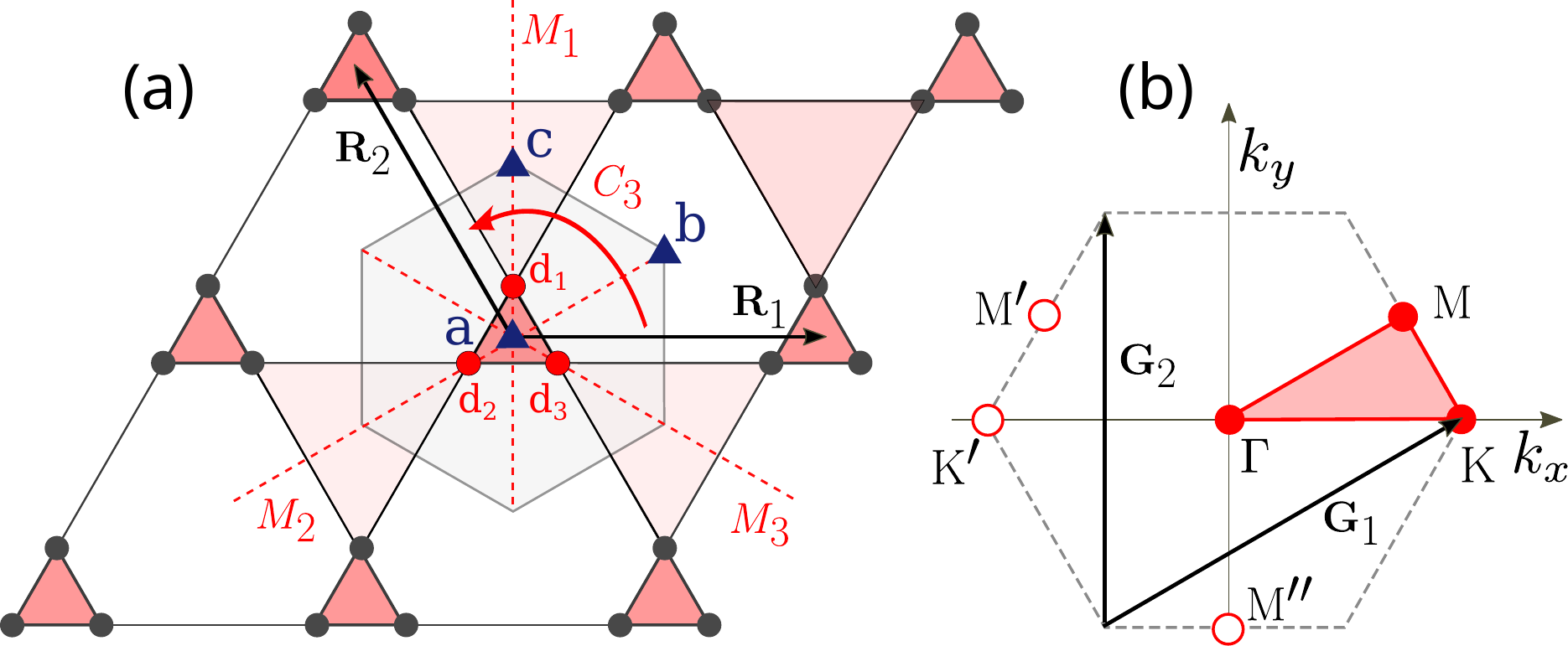}
\caption{\label{fig:lattice} 
(a): Illustration of the breathing kagome lattice. The vectors $\vR_{1,2}$ denote the unit vectors of the underlying triangular Bravais lattice, whose Wigner-Seitz unit cell is the gray hexagon. The three unit-multiplicity Wyckoff positions ``a'', ``b'', and ``c'' are depicted by blue triangles, while the multiplicity-three Wyckoff position ``d'' is denoted by red circles. Note that the Wyckoff positions ``a'' and ``c'' lie at the centroids of the up- and down-pointing triangles, respectively, while the Wyckoff position ``b'' lies at the centroid of a $D_3$-symmetric hexagon. 
(b): The Brillouin zone with the fundamental domain shaded red. Knowledge of the Hamiltonian on the fundamental domain is sufficient to reconstruct it everywhere on the Brillouin zone.}
\end{figure}

\begin{figure}[t]
  \centering
  \includegraphics[width=\columnwidth]{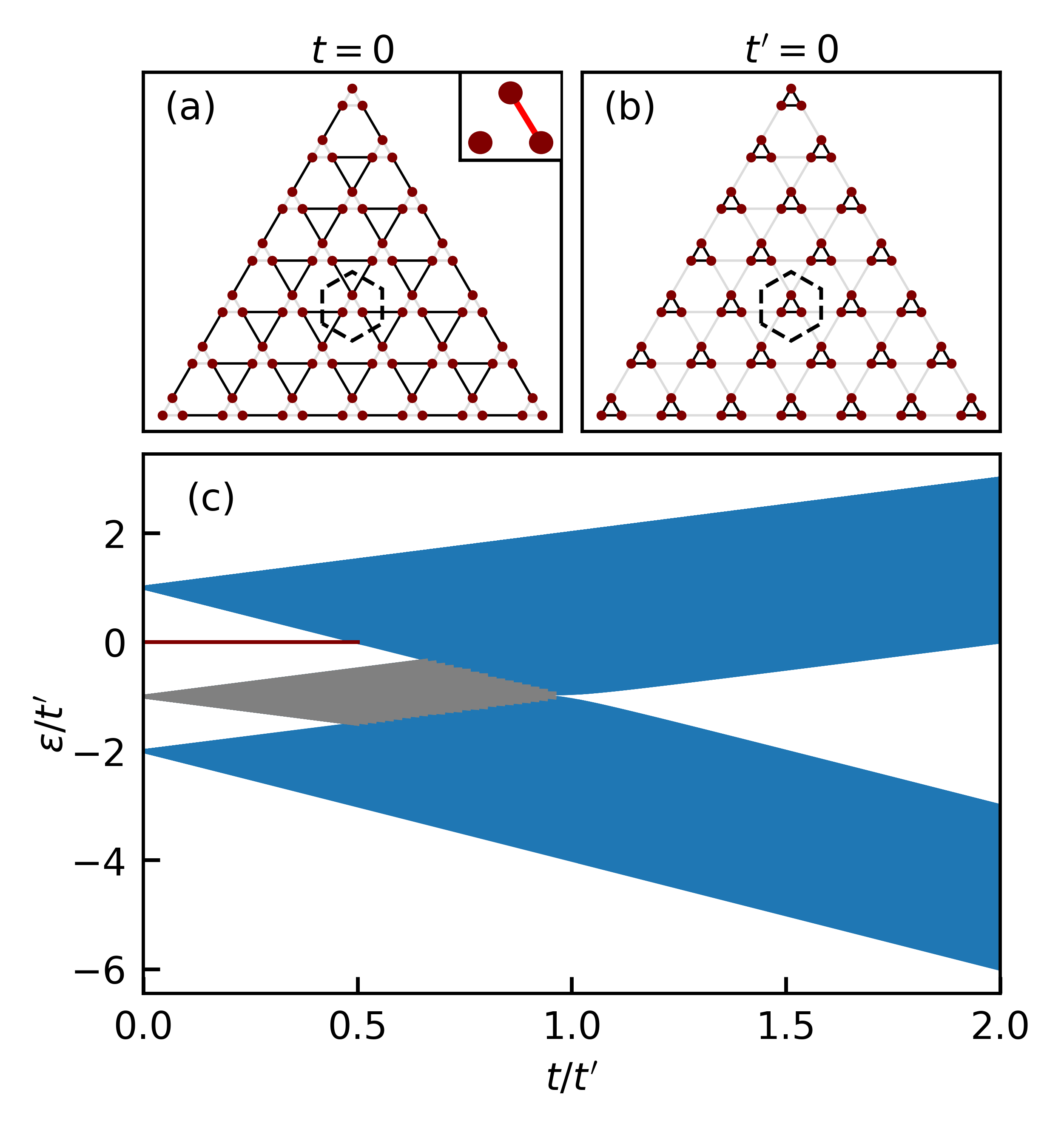}
\caption{\label{fig:spec} 
(a): The breathing kagome lattice in a triangular geometry in the limiting case where intra-unit-cell bonds are set to $t = 0$. The appearance of zero-energy corner states is manifest in this limit. The inset is an example of a triangular unit cell with a single-zero energy state, showing that a single zero-energy state is not anomalous for a tripartite lattice.
(b): Same as (a), but for the limiting case where the inter-unit-cell bonds are set to $t' = 0$.
(c): Spectrum of a large triangular lattice with termination similar to (a) and (b) as a function of $t/t'$ with bulk bands (blue), edge bands (grey), and zero-energy corner states (red). 
}
\end{figure}

{\em The lattice model ---} The breathing kagome lattice is formed by corner-sharing triangles, whereby the up- and down-pointing triangles are inequivalent, as depicted in Fig.~\ref{fig:spec}. It is obtained from a triangular lattice with the point group $p3m1$ consisting of threefold rotation and reflections, which together constitute the threefold dihedral group $\D_3$. The breathing kagome lattice is obtained by only occupying the Wyckoff position ``d'' with multiplicity three, as shown in Fig.~\ref{fig:lattice}(a). The corresponding Brillouin zone is a hexagon with opposite sides identified. The threefold rotation symmetry maps the three distinct $\M$ points onto one another, while the mirror symmetry maps $\K \to \K'$.

The lattice model that purportedly realizes a second order topological phase is a nearest-neighbor hopping model on the breathing kagome lattice, with further imposition of time-reversal symmetry. In the literature an additional ``tripartite symmetry'' is sometimes imposed \cite{kempkes2019robust,ni2019,li2020,li2021}, whereby there only exist hoppings between the three different sublattices. With the tripartite symmetry, the Bloch Hamiltonian then takes the general form 
\begin{equation}
	H(\vk) = 
	\begin{pmatrix}
		0 			& c(\vk) 		& b^\ast(\vk) 	\\ 
		c^\ast(\vk) & 0 			& a(\vk) 		\\ 
		b(\vk) 		& a^\ast(\vk) 	& 0 			\\ 
	\end{pmatrix},  
	\label{eq:hlt_kagome}
\end{equation}
where the lattice symmetries impose further constraints
\begin{equation}
    a(\vk) = a^\ast(M_1\vk) = b(\C_3 \vk) = c(\C_3^2 \vk),
    \label{eq:D3_constr}
\end{equation}
where $\C_3$ and $M_1$ refer to the threefold rotation and mirror operations, respectively. Time-reversal symmetry gives the additional constraint
\begin{equation}
    a(\vk) = a^\ast(-\vk). \label{eq:TRS}
\end{equation}
The explicit form of various symmetry operators in this basis are described in App.~\ref{app:clfn}. Using these relations, the knowledge of either $a(\vk)$ over half the Brillouin zone or that of the trio $a(\vk)$, $b(\vk)$, and $c(\vk)$ over the fundamental domain $\G\M\K$ (which is 1/12 of the full Brillouin zone, see Fig.~\ref{fig:lattice}(b)) is sufficient to specify the Hamiltonian completely.

For the nearest neighbor hopping model with intra- and inter-unit-cell hopping amplitudes $t$ and $t'$, respectively, we have 
\begin{equation}
    a(\vk) = -t - t' \e^{-\I k_1},  
    \label{eq:a_def} 
\end{equation}
with $b(\vk)$ and $c(\vk)$ determined by Eq.~\eqref{eq:D3_constr}. Here, $k_i = \vk\cdot\vR_i$ for $i=1,2$, where $\vR_{1,2}$ are the two lattice vectors that generate the triangular lattice, see Fig.~\ref{fig:lattice}(a).
The bulk spectrum of this system is gapped with a single band below the gap for $\abs{t/t'}<1$. On a finite triangular flake, one additionally gets an edge band and three zero-energy states localized at the corners, which appear for $-1 < t/t' < 1/2$, see Fig.\ \ref{fig:spec}(c) \cite{ezawa2018}. The origin of these corner states can be most easily understood from the two extreme limits $t=0$ and $t'=0$ depicted in Fig.~\ref{fig:spec}(a,b). For the former, we note that while all the bulk and edge sites are coupled to another site, the corner sites are completely decoupled from the bulk and the edge, leading to a zero-energy mode localized at the corner. 

However, the presence of a \emph{single} zero-energy state in a system with tripartite symmetry is not anomalous. A counterexample is provided by the inset of Fig.~\ref{fig:spec}(a). The absence of a spectral symmetry around $\ve = 0$ (see Fig.~\ref{fig:spec}(c)) --- implying an absence of an antisymmetry of $H$ --- further speaks against the identification of the corner modes as topological. We now show by explicit construction that a change of nearest-neighbor hopping amplitudes in the vicinity of the corner can fully remove the zero-energy corner states without violating the crystalline symmetries.

\begin{figure}[t]
\centering
\includegraphics[width=1.05\columnwidth]{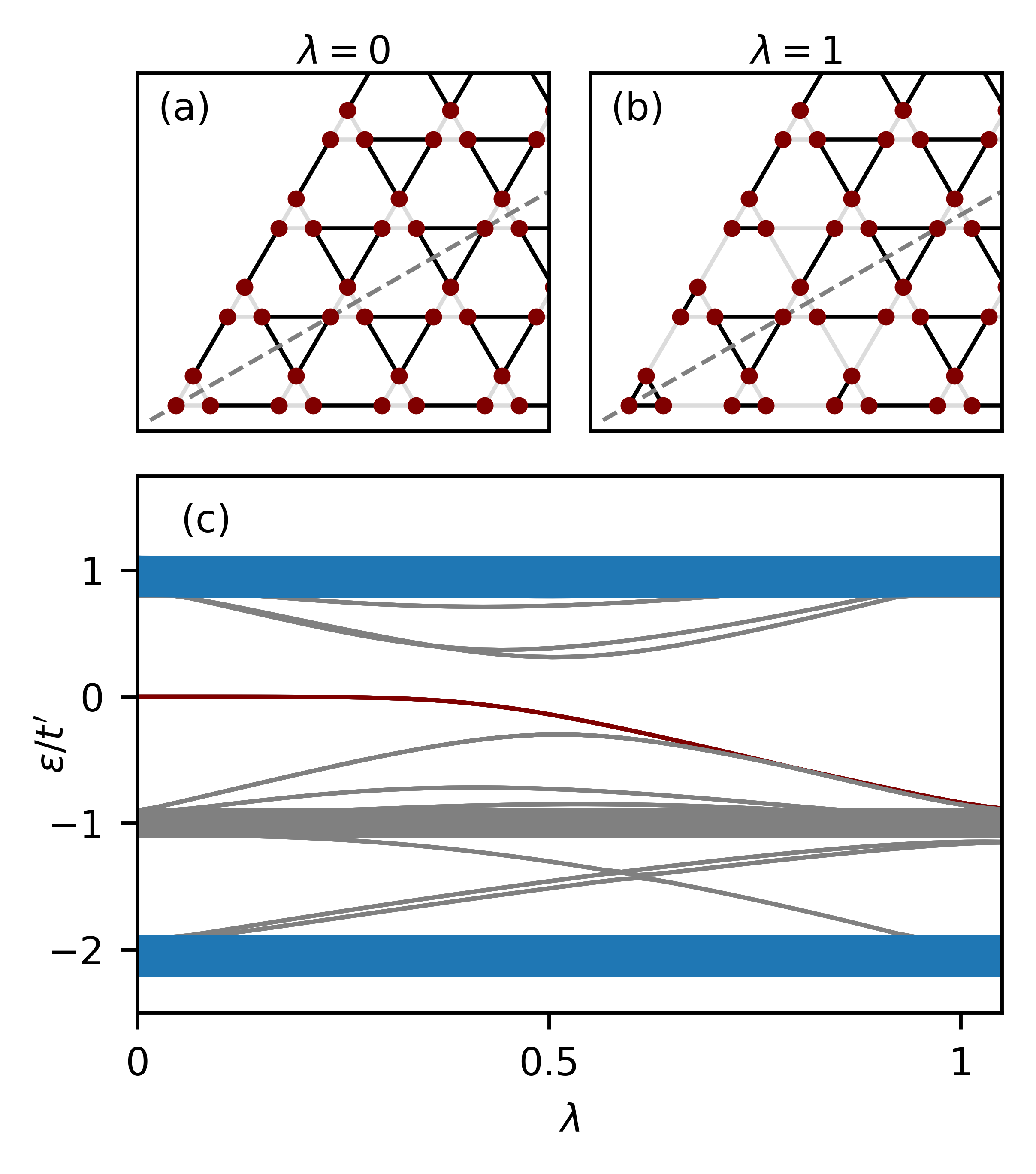}
\caption{\label{fig:interpol_spec} 
(a): Zoom-in of a corner of the breathing kagome lattice of Fig.\ \ref{fig:spec}(a). 
(b): Rearrangement of the hopping strengths within a distance of two unit cells from the corner, so that the zero-energy eigenstate is removed. Bonds in the bulk or at sample edges farther than two unit cells from the corner are the same as in (a). The bond pattern of (b) obeys the mirror symmetry with reflection to the angle bisector of the corner (shown dashed). 
(c) Spectrum with $t=0.1$ and $t'=1$ for bulk bands (blue), edge bands (grey), and zero-energy corner states (red) as a function of the interpolation parameter $\lambda$. }
\vspace{-0.5cm}
\end{figure}

{\em Removing the corner states ---}
We start from the limiting case of $t=0$, for which the corner site is decoupled from the bulk, see Fig.\ \ref{fig:interpol_spec}(a). Figure \ref{fig:interpol_spec}(b) shows a changed pattern of non-zero hopping amplitudes, which differs from the one of Fig.\ \ref{fig:interpol_spec}(a) only within a distance of two unit cells from the corner, while respecting the mirror symmetry with respect to the angle bisector of the corner and the tripartite lattice condition. The threefold rotation symmetry and the two remaining mirror symmetries are not relevant in the vicinity of a corner, because these are already broken by the sample geometry; we can, however, ensure that the overall system respects all lattice symmetries by performing an identical deformation at each corner. The model of Fig.\ \ref{fig:interpol_spec}(b) has no zero-energy state, since it has no isolated lattice sites. It also does not have other localized states with energy outside the (flat) bulk and surface bands. 

To show that we can continuously deform the system of Fig.\ \ref{fig:interpol_spec}(a) into \ref{fig:interpol_spec}(b), we consider a linear interpolation 
\begin{equation}
  H_{\lambda} = (1-\lambda) H + \lambda H'
\end{equation}
between the models $H$ and $H'$ of Figs.\ \ref{fig:interpol_spec}(a) and (b), such that $H_0 = H$ and $H_1 = H'$. For $t=0$, only the energies of $12$ states near the corner change with the interpolation parameter $\lambda$, while the flat bands of edge and bulk states farther away from the corner are unaffected. Moving away from the special point $t=0$, in Figure \ref{fig:interpol_spec}(c) we show the spectrum of $H_{\lambda}$ for hopping amplitudes $t=0.1$ and $t'=1$. In this case, the bulk and edge bands have a finite width and feature extended eigenstates. Upon increasing $\lambda$, the corner mode moves away from zero and merges with the edge band. The edge bands can be further moved out of the bulk gap by way of a perturbation along the entire edge, as we show in App.\ \ref{app:edge_def}. We therefore confirm the conclusion of Refs.~\cite{vanmiert2020,herrera2022} that the breathing kagome does not have second-order topology, but without the need of invoking hopping amplitudes beyond the nearest-neighbor ones present in the original model.

{\em Obstructed atomic limits ---}
Though the breathing kagome lattice with nearest-neighbor hoppings does not have topologically protected anomalous corner states, the models with $|t/t'| < 1$ and $|t/t'| > 1$ are in distinct topological phases. These can be distinguished by a \emph{filling anomaly}, which is the property that, despite having a charge-neutral unit cell, a finite, symmetry-respecting lattice is unable to be charge neutral. For the breathing kagome lattice, the filling anomaly can be characterized by 
\begin{equation}
    \nu = Q\mod 3,
\end{equation}
where $Q$ is the total charge in the ground state (including the positive ions). The trivial phase corresponds to $\nu=0$, while $\nu = \pm1$ correspond to (topologically nontrivial) obstructed atomic limits. 

The filling anomaly can be readily read off from the location of the Wannier centers of the single occupied band, which must be at one of the three highest-symmetry Wyckoff positions. There are three such Wyckoff positions (labelled ``a'', ``b'', and ``c'' in Fig.~\ref{fig:lattice}(a)), located at the center or at the corners of the unit cell (see App.\ \ref{app:clfn} for details). These three Wyckoff positions have different behavior under a threefold rotation around the center of the unit cell: they come with different $\C_3$-eigenvalues $e^{2 \pi i \nu/3}$ at the $\C_3$-symmetric $\K$-point in reciprocal space. This gives $\nu = 0$ for $|t/t'| > 1$ and $\nu = 1$ for $|t/t'| < 1$.

Alternatively, the Wannier centers can be immediately read off in the limiting cases $t/t' = 0$ and $t'/t = 0$, for which the lattice model splits into a set of isolated triangles, bonds at the edge, and sites at the corners. Considering these limits as representatives of their topological phase, one again finds that the two gapped phases with $|t/t'|<1$ and $|t/t'|>1$ exhibit filling anomaly with $\nu=1$ and $\nu=0$, respectively. In the latter case, all Wannier centers are at the center of the unit cell, so that they are cancelled by the charge of the positive ion. In the former case, the Wannier functions are centered at the edges of the unit cell, giving a charge surplus/deficit from unit cells that are cut-off at the edge, if the termination is such that the entire system has $D_3$ symmetry. 

{\em Delicate classification ---}
We finally report an exhaustive classification of all band structures realized by lattice models on the kagome lattice with a single orbital per site and a single occupied band, constrained by threefold rotation, mirror, and time-reversal symmetries but without the restriction to nearest-neighbor hopping. In Apps.\ \ref{app:tripartite} and \ref{app:w/o_tripartite} we derive these classifications with and without an additional tripartite symmetry constraint, where for the former, we additionally fix the Fermi level at $\varepsilon = 0$. In the literature, topology of this type, in which the number of occupied and unoccupied bands is kept fixed, is referred to as {\em delicate} \cite{nelson2021multicellularity,nelson2022,brouwer2023}. From the perspective of classification of topological phases, the tripartite symmetry represents an interesting case, since the bands cannot be flattened without violating the symmetry, so that the relevant target spaces for the Bloch Hamiltonians are not symmetric spaces. 

With tripartite symmetry, we find that there is a single integer-valued topological invariant $n_{\K} \in \ZZ$; without tripartite symmetry, there is a $\ZZ_3 \times \ZZ_2$ classification, with a $\ZZ_3$ invariant $\nu_{\K}$ and a $\ZZ_2$ invariant $\nu_{\Gamma\K}$. (Note, however, that, since the number of bands is kept fixed, the classification does not have a group structure.) In both cases, the topological invariants are defined in terms of the Bloch Hamiltonian on and between the high-symmetry momenta $\G$ and $\K$. The topological invariants are related to the filling-anomaly index $\nu$ as
\begin{align}
  \nu = n_{\K} = \nu_{\K} \mod 3.
\end{align}

The topological invariants of the two nearest-neighbor hopping models on the kagome lattice are $n_{\K} = \nu_{\K} = 0$, $1$ and $\nu_{\G\K} = 0$. These models realize filling anomalies $\nu=0$ and $\nu=1$ only. An example of a model with tripartite symmetry and arbitrary topological invariant $n_{\K}$ is found by setting
\begin{equation}
  a(\vk) = - t'' e^{-\I n_{\K} k_1} \label{eq:achoice}
\end{equation}
in Eq.\ (\ref{eq:hlt_kagome}), with the remaining coefficients $b(\vk)$ and $c(\vk)$ determined by Eq.~\eqref{eq:D3_constr}. Setting $n_{\K} = -1$ results in a model with filling anomaly $\nu=-1$. Note, however, that, despite the simplicity of Eq.\ (\ref{eq:achoice}), this not a model with nearest-neighbor hopping.

{\em Conclusion ---} 
We have shown that the zero-energy corner modes of the nearest-neighbor hopping model on the breathing kagome lattice are not topologically protected and that they can be removed by a local deformation at the corner that only affects nearest-neighbor bonds. Instead, the topology of the breathing kagome lattice is that of an obstructed atomic limit, which exhibits a filling anomaly. Furthermore, we have comprehensively classified all band structures on the kagome lattice with one occupied band and two unoccupied bands and identified which of these have a filling anomaly and fractional corner charges.

The absence of protected zero-energy corner states in hopping models on the kagome lattice is in contrast to the zero-energy corner states of a bona-fide two-dimensional second-order topological insulator, such as the Benalcazar-Bernevig-Hughes (BBH) model \cite{benalcazar2017,benalcazar2017b}. In the BBH model, the bipartite lattice structure imposes a sublattice symmetry that is essential for the protection of the zero-energy corner states against local perturbations. The zero-energy corner modes are robust to arbitrary symmetry-preserving changes of hopping amplitudes in the vicinity of a corner, provided these respect the sublattice symmetry. This robustness also follows from the simple observation that an isolated single zero-energy state is {\em anomalous}: with sublattice symmetry, no lattice built from whole unit cells can host an odd number of zero-energy states.

{\em Acknowledgements. ---} 
We would like to thank Shozab Qasim for helpful discussions. 
This work was supported by the Deutsche Forschungsgemeinschaft (DFG, German Research Foundation) - Project Number 277101999 - CRC TR 183 (projects A02 and A03).

\bibliography{refs}

\clearpage 
\appendix 

\renewcommand\thefigure{S\arabic{figure}}    
\setcounter{figure}{0}    
\renewcommand\thetable{S\arabic{table}}    
\setcounter{table}{0}    

\section{Deformation to a model without edge states in the bulk gap}
\label{app:edge_def}

In the main text, the deformation from the original breathing kagome lattice $H$ to a model $H'$ that differs from $H$ only within a finite distance from the corners was described. The model $H'$ has no zero-energy corner localized states, but still has in-gap states localized along the edges. This is no surprise, since the original model $H$ also has such in-gap edge modes. Here we show that a symmetry-preserving deformation along the crystal edge can be used to completely remove the edge states from the bulk gap.

\begin{figure}[t]
    \centering
    \includegraphics[width=\columnwidth]{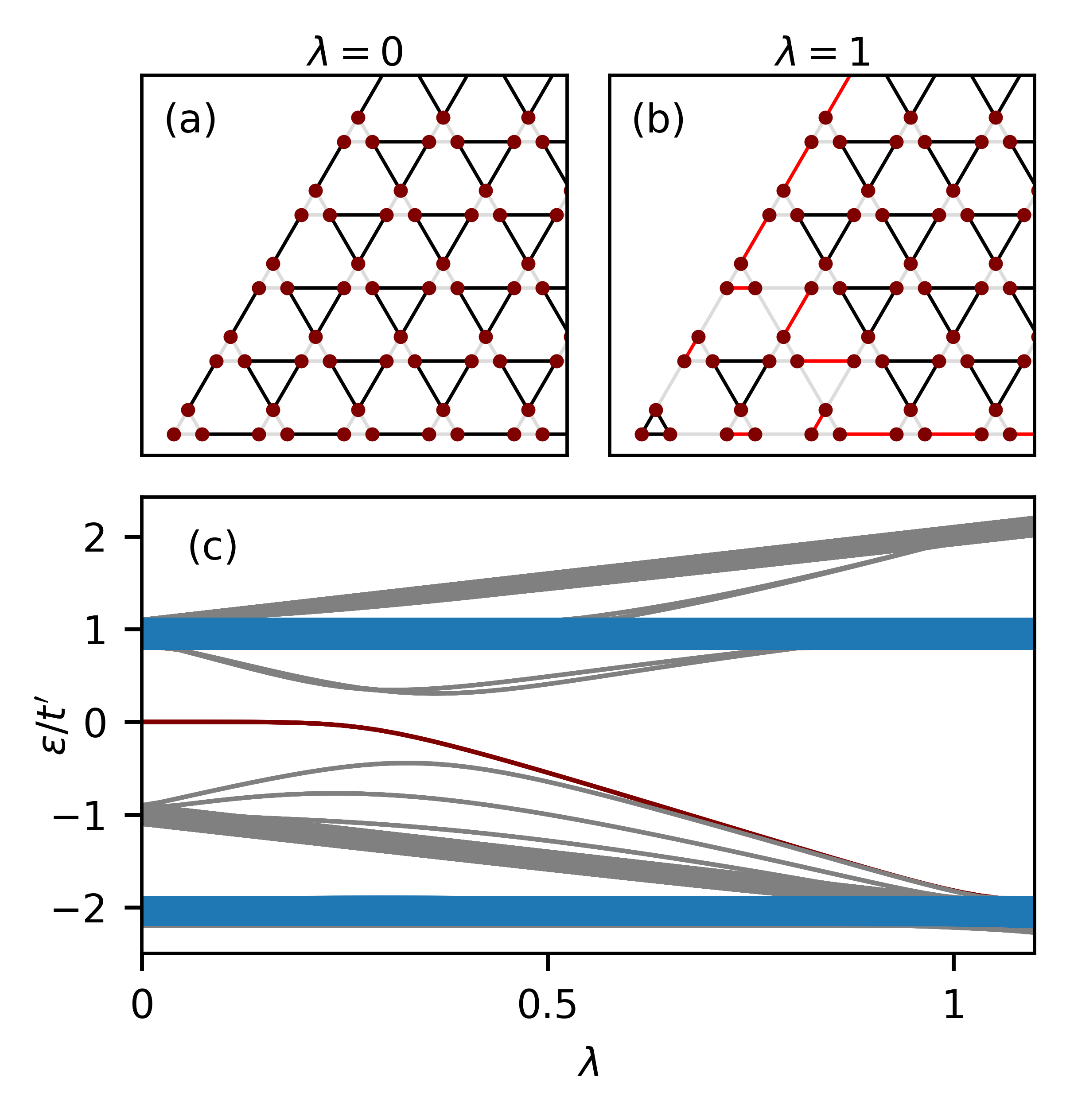}
    \caption{
    (a): Zoom-in of the corner of the breathing kagome lattice at the starting point of the deformation, $\lambda=0$. The gray and black bonds correspond to $t = 0.1$ and $t' = 1$ respectively. This system hosts zero-energy corner states and a band of edge states in the bulk gap.
    (b): Same as (a) at $\lambda=1$. The gray, black and red bonds correspond to hopping strengths $t = 0.1$, $t' = 1$ and $\tau=2$, respectively. Bonds in the bulk farther than two unit cells from the edge are the same as in (a). This system has corner and edge modes removed from the bulk gap.
    (c) Spectrum of bulk bands (blue), edge bands (grey), and zero-energy corner states (red) as a function of the interpolation parameter $\lambda$ with $t=0.1$, $t'=1$, and $\tau = 2$.}
    \label{fig:edge_interpol_spec}
\end{figure}

We deform $H$ to a  model $H''$ which differs from $H'$ only within a finite distance from the \emph{edge}: It differs by the strength of hopping amplitudes $\tau = 2$ on bonds that are depicted in red in \ref{fig:edge_interpol_spec}(b). (The other bonds in Fig.\ \ref{fig:edge_interpol_spec}(b) are the same as for the model $H'$: Grey bonds have hopping $t = 0.1$ and black bonds have hopping $t' = 1$.)
The deformation from $H$ to $H''$ is again performed by way of a linear interpolation 
\begin{equation}
    H_{\lambda} = (1-\lambda) H + \lambda H''. 
\end{equation}
Figures \ref{fig:edge_interpol_spec}(a) and (b) show the patterns of hopping amplitudes corresonding to $H_0 = H$ and $H_1 = H''$, respectively. 
The spectrum of $H_{\lambda}$ is shown in Figure \ref{fig:edge_interpol_spec}(c). By continuous deformation, all corner and edge states are removed from the bulk gap without breaking any symmetries or modifying model properties in the bulk.

\section{Classification of tight-binding models on the breathing kagome}
\label{app:clfn}
The breathing kagome lattice is generated from a triangular Bravais lattice with point group $\D_3$, consisting of threefold rotation and mirror symmetries. In Fig.\ \ref{fig:lattice} of the main text, we make the choice
\begin{equation}
  \vR_{1} = \begin{pmatrix} 1 \\ 0 \end{pmatrix}, \quad 
  \vR_{2} = \frac{1}{2} \begin{pmatrix} -1 \\ \sqrt{3} \end{pmatrix}
\end{equation}
for the two lattice vectors $\vR_1$ and $\vR_2$ that generate the triangular lattice. The relevant Wyckoff positions for this point group are the three special Wyckoff positions ``a'', ``b'', and ``c'' with multiplicity one and a special Wyckoff position ``d'' with multiplicity three, which is on the lines connecting ``a'' and ``c'', see Tab.~\ref{tab:D3_wyc} and Fig.~\ref{fig:lattice}(a). The breathing kagome lattice arises by occupying only the Wyckoff position ``d''. If ``d'' is halfway between ``a'' and ``c'', one recovers the kagome lattice, which has nearest-neighbor bonds of equal length. 

To classify tight-binding models on this lattice, we require that the Bloch Hamiltonian be invariant under threefold rotation and reflection. Explicitly, 
\begin{align}
  H(\vk)
  = \rotn_3^{\rm T} H(\C_3 \vk) \rotn_3 
  = \mirror_1^{\rm T} H(\M_1 \vk) \mirror_1,
\end{align}
whereby in the site basis of the three ``d'' positions (labeled d$_1$, d$_2$, and d$_3$ in Tab.~\ref{tab:D3_wyc} and Fig.~\ref{fig:lattice}), the symmetry operators are explicitly given by 
\begin{equation}
	\rotn_3 = 
	\begin{pmatrix}
		0 & 0 & 1 \\ 
		1 & 0 & 0 \\ 
		0 & 1 & 0 \\ 
	\end{pmatrix}, \qquad 
	\mirror_1 = 
	\begin{pmatrix}
		1 & 0 & 0 \\ 
		0 & 0 & 1 \\ 
		0 & 1 & 0 \\ 
	\end{pmatrix}.
 \label{eq:symm_mat}
\end{equation}
We further impose a (spinless) time-reversal symmetry 
\begin{equation}
    H(\vk) = H^\ast(-\vk).
\end{equation}
These condition imply that it is sufficient to consider the Hamiltonian $H(\vk)$ on the {\em fundamental domain}, the minimal region of the Brillouin zone whose orbit under the reciprocal point group generates the entire Brillouin zone. In the present case, the fundamental domain consists of the triangle $\Gamma\K\M$, as shown in Fig.\ \ref{fig:lattice}(b) of the main text. The little groups for various high-symmetry points and lines constituting the fundamental domain are summarized in Tab.\ \ref{tab:target_spaces}.


\subsection{Classification with tripartite symmetry}
\label{app:tripartite} 
The tripartite symmetry implies that the lattice can be divided into three sublattices such that no bonds are between sites on the same sublattice, or equivalently, that the Bloch Hamiltonian satisfies \cite{ni2019}
\begin{equation}
  H(\vk) + \Omega H(\vk) \Omega^\dg + \Omega^2 H(\vk) (\Omega^2)^\dg = 0, 
\end{equation}
whereby $\Omega = \mbox{diag}\,(1, \e^{2\pi\I/3}, \e^{4\pi\I/3})$. This condition implies that the Bloch Hamiltonian $H(\vk)$ has the form of Eq.\ (\ref{eq:hlt_kagome}), which contains the three complex functions $a(\vk)$, $b(\vk)$, and $c(\vk)$. We begin by parametrizing these as
\begin{align}
  a &= - \abs{a} \e^{\I (\psi/3 + \theta)},\nonumber \\
  b &= - \abs{b} \e^{\I (\psi/3 + \eta)},\nonumber \\
  c &= - \abs{c} \e^{\I (\psi/3 - \theta - \eta)}, 
  \label{eq:abc_param}
\end{align}
where $\theta$, $\eta$, $\psi \in [-\pi, \pi)$ and we have suppressed their explicit $\vk$-dependence to avoid notational clutter. The eigenvalue problem for $H$ can be solved exactly using its tracelessness as well as 
\begin{equation}
    \tr H^2 = 2\left( a^2 + b^2 + c^2 \right), \quad 
    \det H = 2 \mathrm{Re} (abc).
\end{equation}
The three eigenvalues of the Hamiltonian are given by 
\begin{equation}
    \ve_n = 2\sqrt{\frac{a^2 + b^2 + c^2}3 } \sin\left( \phi + \frac{2n\pi}3 \right) 
\end{equation}
where $n = -1,0,1$ and 
\begin{equation}
  \phi = \frac{1}{3} \sin^{-1} \left( \frac{\abs{abc} \cos\psi}{\left( a^2 + b^2 + c^2 \right)^{3/2}} \right).  
\end{equation}
We are interested in Bloch Hamiltonians with one negative eigenvalue and two positive eigenvalues, which is the case if $a,b,c$ are nonzero and 
\begin{equation}
    \phi \in \left( 0, \frac\pi6 \right] \implies 
    \psi \in \left( -\frac\pi2, \frac\pi2 \right). 
\end{equation}
For nonzero $a(\vk)$, $b(\vk)$, and $c(\vk)$, the parametrization (\ref{eq:abc_param}) is unique and continuous over the Brillouin zone, provided $\theta(\vk)$ and $\eta(\vk)$ satisfy periodic boundary conditions. 

\begin{table}
	\centering 
	\setlength\tabcolsep{20pt}
	\begin{tabular}{ccl}
		\hline
		$\alpha$ & $G_\alpha$  & position $\vr_{\alpha}$ \\
		\hline 
		a & \multirow{3}{*}{$\D_3$} & $\nullv$ \\ 
		b & & $(\vR_1 + 2\vR_2)/3$ \\ 
		c & & $(2\vR_1 + \vR_2)/3$ \\ 
		\hline
        d$_1$ & & $\wpos (2\vR_1 + \vR_2)$ \\
        d$_2$ & $\intg_2$ & $-\wpos(\vR_1 + 2\vR_2)$ \\
        d$_3$ & & $\wpos(\vR_1 - \vR_2)$ \\
		\hline
	\end{tabular}
	\caption{The relevant Wyckoff positions $\alpha$, their position $\vr_\alpha$ within the unit cell, and the corresponding little groups $G_\alpha$ for a two-dimensional $\D_3$-symmetric lattice. The coefficient $\wpos\in(0,1/3)$ indicates the position of the Wyckoff positions d along the line ac. The choice $\wpos = 1/6$ corresponds to the kagome lattice. The breathing kagome lattice has $0 < \wpos < 1/6$ or $1/6 < \wpos < 1/3$.}
	\label{tab:D3_wyc} 
\end{table}

\begin{table*}
	\centering 
	\setlength\tabcolsep{15pt}
	\begin{tabular}{ccccc}
		\hline
		$\omega$ & Little group $G_\omega$ & Generators of $G_\omega$ & Constraints on $(a,b,c)$ & Constraints on $(\psi, \theta, \eta)$
                \\ 
		\hline 
		$\G$ & $\D_3 \times \intg_2$  & $\rotn_3$, $\mirror_1$, $\trev$ & $a = b = c \in \real$ & $\psi = \theta = \eta = 0$ 
                \\ 
		$\K$ & $\C_3 \times \intg_2$  & $\Lambda_2 \rotn_3, \mirror_1\trev$ & $a = b = c$ & $\theta = \eta \in \{0, \pm2\pi/3\}$ 
                \\ 
		$\M$ & $\intg_2 \times \intg_2$  & $\mirror_3, \Lambda_1 \trev$ & $a = b; \; a,b,c \in \real$ & $\psi = 0$, $\theta = \eta \in \{0,\pi\}$ 
                \\ \hline
		$\G\M$ & $\intg_2$  & $\mirror_3$ & $a = b^\ast; \; c \in \real$ & $\psi = 0$, $\theta = -\eta$ 
                \\ 
		$\G\K$ & $\intg_2$  & $\mirror_1\trev$ & $b = c$ & $\theta = -2\eta$ 
                \\ 
		$\M\K$ & $\intg_2$  & $\Lambda_1\mirror_3\trev$ & $a = b$ & $\theta = \eta$ 
                \\ 
		\hline
	\end{tabular}
	\caption{Little groups and constraints on the coefficients $a$, $b$, and $c$ for high-symmetry points and lines $\omega$ in the fundamental domain. The parameters $a,b,c\in\cmplx$ unless otherwise specified in individual cases. The operations $\Lambda_1$ and $\Lambda_2$ indicate translation by a reciprocal-lattice vector. Note that the constraints on the angles $(\psi, \theta, \eta)$ are defined $\mod 2\pi$.}
	\label{tab:target_spaces} 
\end{table*}

At high-symmetry points and lines, the Hamiltonian must commute with the generators of the corresponding little group, which implies further constraints on $(a,b,c)$, as listed in Tab.~\ref{tab:target_spaces}. We now derive the resulting constraints on $\psi$, $\theta$ and $\phi$. At the $\Gamma$-point, one must have $a(\vk_{\G}) = b(\vk_{\G}) = c(\vk_{\G}) \in \real$ (see Tab.\ \ref{tab:target_spaces}). This implies that 
\begin{equation}
    \psi(\vk_\G) = \theta(\vk_\G) = \eta(\vk_\G) = 0. 
\end{equation}
Here, the vanishing of $\psi$ follows from 
\[
    \e^{\I\psi(\vk_\G)} 
    = -\frac{a(\vk_\G)b(\vk_\G)c(\vk_\G)}{\abs{a(\vk_\G)b(\vk_\G)c(\vk_\G)}}  
    = -\left( \sgn{a(\vk_\G)} \right)^3, 
\]
where $\psi(\vk_\G) \in (-\pi/2,\pi/2)$ and the right hand side equals $\pm1$. This along with the reality of $a,b,c$ further implies that $\theta(\vk_\G), \eta(\vk_\G) \in \{0,\pi\}$, while from their equality, it follows that 
\begin{equation}
    \theta(\vk_\G) = \eta(\vk_\G) = -\theta(\vk_\G)-\eta(\vk_{\Gamma}) \mod 2 \pi,
    \label{eq:theta_eta_Gamma}    
\end{equation}
which implies the vanishing of $\theta(\vk_\G)$ and $\eta(\vk_\G)$. At the $\K$-point, we require that $a(\vk_{\K}) = b(\vk_{\K}) = c(\vk_{\K})$. This yields a condition identical to Eq.~\eqref{eq:theta_eta_Gamma}, which implies that $3\theta(\vk_\K) = 0 \mod 2 \pi$. Thus, 
\begin{equation}
    \theta(\vk_\K) = \eta(\vk_\K) \in \left\{ 0, \pm\frac{2\pi}3 \right\}
\end{equation}
with no constraint on $\psi(\vk_\K)$. Finally, at the $\M$-point, $a(\vk_\M),b(\vk_\M),c(\vk_\M) \in \real$  and $a(\vk_\M) = b(\vk_{\M})$, so that 
\begin{equation}
    \psi(\vk_\M) = 0, \quad \theta(\vk_\M) = \eta(\vk_\M) \in \{0,\pi\}. 
\end{equation}
Along the high-symmetry line $\G\M$, the symmetries require that $a(\vk) = b^\ast(\vk)$ and $c(\vk)\in\real$, so that 
\begin{equation}
    \psi(\vk) = 0, \quad \theta(\vk) = -\eta(\vk) \mod 2\pi, 
\end{equation}
where the vanishing of $\psi$ follows from 
\[
    \e^{\I\psi(\vk)} 
    = -\frac{a(\vk)b(\vk)c(\vk)}{\abs{a(\vk)b(\vk)c(\vk)}}  
    = -\sgn{c(\vk)} \in \{\pm1\}. 
\]
Along $\M\K$, we require $a(\vk) = b(\vk)$, so that 
\begin{equation}
    \theta(\vk) = \eta(\vk) \mod 2\pi, 
\end{equation}
while along $\G\K$, $b(\vk) = c(\vk)$, so that 
\begin{equation}
    \theta(\vk) = -2\eta(\vk) \mod 2\pi. 
\end{equation}
There are no symmetry constraints on $\psi(\vk)$ at $\K$ and along $\G\K$ and $\M\K$.

Since the phase angle $\psi(\vk)$ and the absolute values $|a(\vk)|$, $|b(\vk)|$, and $|c(\vk)|$ take values in sets with trivial topology, only the phase angles $\theta(\vk)$ and $\eta(\vk)$ are of relevance for the topology of $H(\vk)$. As the fundamental domain is simply connected and the parametrization of Eq.\ (\ref{eq:abc_param}) is unique, one may continuously extend the phase angles $\theta(\vk)$ and $\eta(\vk)$ to the entire fundamental domain as real-valued phase angles $\theta(\vk)$, $\eta(\vk) \in \real$.
For definiteness, we choose the phase angles at $\Gamma$ as
\begin{equation}
  \theta(\vk_{\Gamma}) = \eta(\vk_{\Gamma}) = 0,
\end{equation}
Along $\Gamma\M$, we must have $\theta(\vk) = -\eta(\vk) \mod 2 \pi$. Combining this with the constraint at the $\M$-point, we must have 
\begin{align}
  \theta(\vk_{\M}) = \pi n_{\M},\quad
  \eta(\vk_{\M}) = -\pi n_{\M},
\end{align}
with $n_{\M} \in\intg$. Similarly, combining $\theta(\vk) = -2 \eta(\vk) \mod 2 \pi$ along $\G\K$ with the constraint on the $\K$-point, we get 
\begin{align}
  \theta(\vk_{\K}) = -\frac{4 \pi}{3} n_{\K},\quad
  \eta(\vk_{\K}) = \frac{2 \pi}{3} n_{\K},
\end{align}
with $n_{\K} \in\intg$. 
Finally, the condition $\theta(\vk) = \eta(\vk) \mod 2 \pi$ along $\M\K$ implies that 
\begin{equation}
    \theta(\vk_\M) - \theta(\vk_\K) = \eta(\vk_\M) - \eta(\vk_\K),  
\end{equation}
which gives the additional constraint
\begin{equation}
  n_{\M} = n_{\K}.
\end{equation}
We conclude that Hamiltonians $H(\vk)$ that obey the $\D_3$ point group symmetry, time-reversal symmetry, and the tripartite condition, and with one negative eigenvalue and two positive eigenvalues, have a $\ZZ$ classification, with the topological invariant $n_{\K}$.

\subsection{Classification without tripartite symmetry}
\label{app:w/o_tripartite} 
In the absence of tripartite symmetry, we use the homotopic classification scheme of Ref.~\cite{brouwer2023}. This approach requires the interpretation of the $d$-dimensional fundamental domain as a CW-complex, which can be decomposed onto a set of $p$-cells with $0\leq p\leq d$. For each $p$-cell $\omega$, the little group --- a subset of the point group that leaves $\omega$ invariant --- imposes constraints on the Bloch Hamiltonian, and the space of allowed Hamiltonians on $\omega$ constitutes the \emph{target space} ${\cal H}_{\omega}$. The classification starts from 0-cells and proceeds up in dimension. For the classification over $p$-cells (termed ``level-$p$''), one first determines the {\em parent classification group} $\pi_p({\cal H}_{\omega})$, which gives the topological equivalence classes on $p$-cells with $H(\vk)$ equal to a fixed reference Hamiltonian compatible with the topology at levels $< p$ at the boundary of the $p$-cells. In a second step, topological classes in the parent classification that can be related to each other by a continuous deformation at lower levels are identified and classes that are incompatible with a gapped Hamiltonian at higher levels are eliminated. The level-$0$ classification is identical to that obtained in the framework of {\em topological quantum chemistry} \cite{bradlyn2017}; the higher-level classifications give a further refinement of the topological classification.

For the breathing kagome lattice, we decompose the fundamental domain into three $0$-cells ($\G$,$\M$, and $\K$), three $1$-cells ($\G\M$, $\M\K$, and $\G\K$), and one $2$-cell ($\G\K\M$). Below, we first determine the target space for each cell in the fundamental domain and then we perform the classification procedure, starting from the $0$-cells and working our way upwards.

\newcommand\invminus{\phantom{-}}
\begin{table}[b]
	\centering
	\setlength\tabcolsep{8pt}
	\begin{tabular}{cccc}
		\hline
		Irrep label    & $\{\id\}$   & $\{\rotn_3, \rotn_3^2\}$   & $\{\mirror_1, \mirror_2, \mirror_3 \}$ \\
		\hline     
		$A_1$   & 1   & $\invminus 1$  & $\invminus 1$   \\ 
		$A_2$   & 1   & $\invminus 1$  & $-1$   \\ 
		$B$   & 2   & $-1$  & $\invminus 0$  \\ 
		\hline
	\end{tabular}
	\caption{The character table of $\D_3$. The character corresponding to $\{\id\}$ is equal to the dimensionality of the representation.} 
	\label{tab:D3_chr}
\end{table}

{\em Target spaces.---} At a generic $\vk$ in the interior of the fundamental domain, $H(\vk)$ is an arbitrary Hermitian matrix with one occupied and two unoccupied bands, so that the target space
\begin{equation}
  {\cal H} = \Gr_{\cmplx}(1,2).
\end{equation}
We next compute the target space ${\cal H}_{\omega} \subseteq {\cal H}$ for individual $p$-cells $\omega$ with $p=0$, $1$. 

At the $\G$-point, since the little group is $\D_3 \times \intg_2$, with the $\intg_2$ corresponding to time-reversal symmetry, the Hamiltonian is block diagonal with individual blocks corresponding to irreps of $\D_3$, see Tab.~\ref{tab:D3_chr}. Time-reversal symmetry does not impose any further constraints, since all irreps of $\D_3$ are real. The representation matrices of Eq.~\eqref{eq:symm_mat} form a reducible representation of $\D_3$, which can be decomposed into irreps as $A_1 \oplus B$, where $A_1$ denotes the trivial irrep and $B$ the two-dimensional irrep of $\D_3$ (see Tab.~\ref{tab:D3_chr}). Since there is only one occupied band, it must transform under $A_1$, whereas the unoccupied band transforms under $B$. Furthermore, time-reversal symmetry requires that $H(\vk_{\Gamma})$ be real-valued. Thus
\begin{align}
  {\cal H}_{\Gamma} =&\, \Gr_{\real}(1,0) \times \Gr_{\real}(0,1) = \{ 1 \},
\end{align}
where $\{ 1 \}$ denotes the set with a single element. 

For the high-symmetry line $\G\M$, the little group $\intg_2 \subset \D_3$ consists only of the mirror symmetry $\mirror_2$. Since the irrep $A_1$ has a positive mirror parity, while $B_1$ consists of a positive and a negative mirror parity state, we get 
\begin{align}
  \hsp_{\G\M} =&\, \Gr_\cmplx(1,1) \times \Gr_\cmplx(0,1) = \Gr_\cmplx(1,1).
\end{align}
Consequently, at the $\M$-point, which is further invariant under time-reversal, we get 
\begin{equation}
	\hsp_{\M} = \Gr_\real(1,1),
\end{equation}
where the reality condition follows from time-reversal symmetry. 

At the $\K$-point, the little group $\C_3 \times \intg_2$ is generated by threefold rotation and time-reversal (in combination with a mirror operation). The latter forces the Hamiltonian to be real, while the former requires that it be a diagonal matrix, since the representation matrix $\rotn_3$ of Eq.~\eqref{eq:symm_mat} is a direct sum of the three irreps of $\C_3$, which are all one-dimensional. The Hamiltonian is thus determined completely by the specification of the $\C_3$-eigenvalue of the occupied state, so that
\begin{equation}
	\hsp_\K \cong \ZZ_3,
\end{equation}
where the three elements of $\ZZ_3$ refer to the three $\C_3$-eigenvalues. Finally, the little groups of $\G\K$ and $\M\K$ are generated by a time-reversal symmetry (again in combination with a mirror operation), so that 
\begin{align}
	\hsp_{\G\K} = \hsp_{\M\K} = \Gr_\real(1,2). 
\end{align}

{\em Homotopic classification.---} 
The only nontrivial level-$0$ classification group $\hom0{\hsp_\omega}$ is at $\K$, where $\hom0{\hsp_\K} = \intg_3$. The corresponding topological invariant $\nu_{\K}$ is the $\C_3$-eigenvalue $e^{2 \pi i \nu_{\K}/3}$ of the occupied state at the $\K$-point.

The parent classification spaces for the three $1$-cells are 
\begin{align}
	\hom1{\hsp_{\G\M}} = 0, \quad 
    \hom1{\hsp_{\G\K}} = \hom1{\hsp_{\M\K}} \cong \intg_2,   
\end{align}
where the invariants on $\G\K$ and $\M\K$ are given by the determinant of the Wilson loop of the occupied band along $\G\K$ and $\M\K$, respectively. There is no compatibility constraint, because $\hom1{\hsp} = 0$. The parent invariant associated with $\M\K$ is not robust to boundary deformations at $\M$, since $\hom1{\hsp_{\M}} \cong \ZZ$, whereby odd elements change the sign of the determinant of the Wilson loop. Hence, at level-$1$ only the invariant $\nu_{\G\K} \in \ZZ_2$ remains. 

There are no topological invariants at level-$2$. To see this we note that the level-$2$ parent classification space is $\hom2{\hsp} \cong \intg$, whereby the invariant is the Chern number. However, the Chern number is not a robust invariant for this symmetry class, since $\hom2{\hsp_{\G\M}} \cong \intg$, so that deformations along the high-symmetry line $\G\M$ can change the Chern number defined on the 2-cell $\G\K\M$ by an arbitrary integer.

Summarizing, we find a homotopic classification with one $\ZZ_3$ invariant $\nu_{\K}$, related to the $C_3$-eigenvalue of the occupied band at $\K$, and one $\ZZ_2$-invariant $\nu_{\G\K}$, corresponding to the determinant of the Wilson loop of the occupied band along $\G\K$.

{\em Relation to classification with tripartite symmetry.---} One may ask, which of these topological equivalence classes a Bloch Hamiltonian with tripartite symmetry and $\ZZ$-valued topological invariant $n_{\K}$ is in (see the discussion of the previous Section). This question can be easily answered by explicit calculation for the realizations given in Eq.\ (\ref{eq:achoice}), for which one finds that
\begin{align}
  \nu_{\K} = n_{\K} \mod 3, \qquad
  \nu_{\G\K} = 0.
\end{align}

\end{document}